\documentclass[aps,twocolumn]{revtex4-2}

\usepackage{amsfonts,amsmath,bm,amssymb,revsymb,color}
\usepackage{graphicx}
\usepackage[makeroom]{cancel}

\usepackage[unicode=true]{hyperref}

\definecolor{bostonuniversityred}{rgb}{0.8, 0.0, 0.0}
\definecolor{dukeblue}{rgb}{0.0, 0.0, 0.61}
\definecolor{ao(english)}{rgb}{0.0, 0.5, 0.0}
\definecolor{darkmagenta}{rgb}{0.55, 0.0, 0.55}
\definecolor{armygreen}{rgb}{0.29, 0.33, 0.13}
\definecolor{coquelicot}{rgb}{1.0, 0.22, 0.0}
\definecolor{fucsiak}{rgb}{0.4, 0.08, 0.4}
\definecolor{airforceblue}{rgb}{0.36, 0.54, 0.66}

\begin{document}

\title{Mixing in two-dimensional shear flow with smooth fluctuations}
\author{Nikolay A. Ivchenko}
\email{ivchenko@itp.ac.ru}
\author{Vladimir V. Lebedev}
\email{lebede@itp.ac.ru}
\author{Sergey S. Vergeles}
\email{ssver@itp.ac.ru}

\affiliation{Landau Institute for Theoretical Physics,
Russian Academy of Sciences,\\1-A Akademika Semenova av.,
142432 Chernogolovka, Russia}
\affiliation{National Research University Higher School of
Economics, Faculty of Physics,\\Myasnitskaya 20, 101000
Moscow, Russia}

\begin{abstract}
Chaotic variations in flow speed up mixing of scalar fields via intensified stirring. This paper addresses the statistical properties of a passive scalar field mixing in a regular shear flow with random fluctuations against its background. We consider two-dimensional flow with shear component dominating over smooth fluctuations. Such flow is supposed to model passive scalar mixing e.g. inside a large-scale coherent vortex forming in two-dimensional turbulence or in elastic turbulence in a micro-channel. We examine both the decaying case and the case of the continuous  forcing of the scalar variances. In both cases dynamics possesses strong intermittency, that can be characterized via the single-point moments and correlation functions calculated in our work. We present general qualitative properties of pair correlation function as well as certain quantitative results obtained in the framework of the model with fluctuations that are short correlated in time.
\end{abstract}

\maketitle

\section{Introduction}

Mixing is a process of homogenization of a scalar field in fluids, such as temperature or a concentration of impurities, accelerated by means of advection. The acceleration is especially effective in chaotic flows which are characterized by irregular fluctuations of the flow velocity in space and/or time. This random part speeds up stirring process for the scalar, that causes stretching of its blobs into lamellae with their subsequent folding. The thinning of lamella in its transverse direction initiates the molecular diffusion that finalizes the mixing. The mixing process has non-trivial statistical properties, see, for example \cite{falkovich2001particles,villermaux2019mixing}, and a problem of passive scalar considers the limit when the field's back reaction to the flow is negligible.

After ascertainment of the statistical properties of developed isotropic three-dimensional \cite{kolmogorov1941equations} and two-dimensional \cite{kraichnan1967inertial} turbulence, the theory of mixing in a statistically isotropic flow within inertial interval \cite{obukhov1949structure,corrsin1951spectrum} and at scales below the Kolmogorov (viscous) one \cite{batchelor1959small} became its natural development,  see also reviews \cite{donzis2010batchelor,sreenivasan2019turbulent}. The isotropic turbulence is an idealized model of the small-scale pulsations imposing on time-averaged large-scale flow component. However, the gradient of the mean flow entails anisotropy which varies statistical properties both of the pulsations and of the mixing process.

The degree of the variation depends on the ratio between gradients magnitudes of the mean flow and the turbulent part. In case of the near-wall turbulence \cite{prandtl1935mechanics,jimenez2013near} in three-dimensional flow they are of same order. Along with the Kolmogorov scaling for velocity lasting at the small scales within inertial range~\cite{smits2011high}, anisotropy of flow spreads down there as well. It produces so-called ramp-cliff structures \cite{sreenivasan2019turbulent} and anisotropy in scalar gradient \cite{germaine2018persistence} in the case of Schmidt number $\mathrm{Sc}=\nu/\kappa\sim 1$, where $\nu$ is the kinematic viscosity of the fluid and $\kappa$ is the diffusion coefficient of the scalar. Another case, the limit where mean flow is prevailing over turbulent pulsations, takes place under some conditions. Axisymmetric vortical flow and near-wall flow are the simplest forms in geometrical sense. In both velocity gradient around a Lagrangian trajectory established by the mean flow remains unchanged thus forming a shear flow. An example of the first type is large-scale coherent vortex emerging in two-dimensional turbulence due to the inverse energy cascade~\cite{xia2009spectrally,laurie2014universal, kolokolov2016structure, kolokolov2016velocity} that motivated the current study. An elastic turbulent flow of polymer solution in micro-channel \cite{groisman2001elastic} is of the second type. For all of them velocity energy spectra are steep enough~\cite{connaughton2007condensation,groisman2000elastic,steinberg2020elastic}, so one can assume the fluctuations on the background of the shear flow are large-scale as well and thus are smooth. Suppression of small-scale turbulent pulsations means that the effective Schmidt number $\mathrm{Sc}$ is increased. Indeed, the Kolomogorov scale should be replaced by the flow scale $R$, so the Schmidt number $\mathrm{Sc}\sim R^2/r_\kappa^2$ \cite{donzis2010batchelor}, where the Batchelor (diffusion) scale $r_\kappa\sim\sqrt{\kappa/\lambdabar}$ and $\lambdabar$ is the Lyapunov exponent of the flow. The same assumption should be applicable to describe mixing in laminar vortex flow of Newtonian fluid \cite{speetjens2021lagrangian,liu2014application}, where the source of the flow fluctuations are imperfect boundary conditions as well as deviations of force driving the flow.

In this work, we provide an analytical study of a passive scalar field $\vartheta$ mixing in a smooth velocity field with strong static shear component and relatively weak fluctuations at large Schmidt number. We consider either the decay of the passive scalar or its continuous forcing. For the decay problem, one starts with certain initial distribution of the passive scalar and examines the evolution of its statistical characteristics. Experimentally, decay of the passive scalar was observed in channels \cite{feng2005investigation}, micro-channels \cite{burghelea2004chaotic} and soap films \cite{amarouchene2004batchelor,jun2010mixing}, where the passive scalar supply is organized at the input to the flow. The decay problem for spatially smooth velocity field with isotropic statistics of turbulent fluctuations in absence of the mean flow was treated analytically in Refs.~\onlinecite{son1999turbulent,balkovsky1999universal}, where moments of the passive scalar were considered. The examination was extended to high-order correlation functions in Ref.~\onlinecite{vergeles2006spatial}. The case of scalar mixing in constant shear flow, which makes the advection deterministic, was considered in Ref.~\onlinecite{souzy2018mixing}. The case when a random component is imposed on the shear flow was considered numerically in Ref.~\onlinecite{celani2005shear} within Kraichnan model. The continuous statistically homogeneous in time stochastic forcing of passive scalar leads to a statistically steady state providing statistics of the random flow is homogeneous in time as well \cite{sreenivasan2019turbulent,donzis2010batchelor}. In large Schmidt number limit, passive scalar cascade develops the Batchelor spectrum in a wide range from the Kolmogorov scale down to the Batchelor scale~\cite{obukhov1949structure,corrsin1951spectrum,batchelor1959small}.

The general picture of passive scalar evolution can be considered in terms of a separate blob. The molecular diffusion effects can be neglected at scales larger than the Batchelor length, where scalar mixing reduces to its advection by the flow. In a stationary shear flow, any vector $\bm\ell$ connecting two close Lagrangian trajectories grows linearly as time goes, aligning along the streamlines. However, random component of the flow causes the tumbling processes, when the direction of $\bm\ell$ is inverted \cite{chertkov2005polymers,turitsyn2007polymer}, so it deviates from the streamlines at time average. As a result, flow's random component enables blob to stretch exponentially in time in one direction and yet to shrink in transverse direction, permanently experiencing tumblings. In experiment, tumblings can be visualized by observing the polymer elongations, see Refs.~\onlinecite{smith1999single,liu2010stretching}. Such processes of passing through unstable stationary point by virtue of fluctuations takes place in various non-equilibrium physical systems \cite{vsiler2018diffusing}. The diffusion effects are switched on when the lateral size of the blob is diminished down to the Batchelor scale. Then the lateral size of the blob is stabilized at this scale whereas the longitudinal size of the blob continues to grow exponentially.This means dissolution of the blob via mixing due to the concentration inside it is inversely proportional to the its area.

The paper is organized as follows. In Section \ref{sec:general} we discuss the general properties of passive scalar dynamics and study statistics of Lagrangian trajectories that describe mixing without diffusion. Some analytical results are obtained in the framework of the model where the random flow is short correlated in time. After that we include diffusion effects into our consideration and examine moments and correlation functions of the passive scalar. The decay problem is analyzed in Section \ref{sec:decay}, where the passive scalar evolution starts from initial distribution in a form of axially symmetric blobs' ensemble. The key component in the investigation of the passive scalar statistics is averaging over the flow statistics. Since the random flow is assumed to have the correlation length much larger than the sizes of the blobs, it coherently influences a lot of blobs. This leads to strongly non-Gaussian statistical properties of the passive scalar that are studied in present work. The continuous forcing of scalar is considered in Section~\ref{sec:continuous}, it is realized via bringing new statistically independent blobs into the system by an external source in our model, after that each evolves as in the decay case. Part of results about the single-point moments was presented in \cite{ivchenko2023statistics}. In the present work we choose expedient technique, which includes rescaling in the streamwise direction, that reformulates the problem and enables to compare it directly to the isotropic turbulence case. Moreover, here we analyse spatial correlation functions dependency at different points. Some technical details are presented in Appendices.

\section{General relations}
\label{sec:general}

In this Section we introduce basic relations required to examine passive scalar statistics. We consider the scalar field $\vartheta(t,\bm r)$ carried by a fluid flow while being diffused and supplied by an external source (pumping). The equation governing passive scalar dynamics is
\begin{equation}
\label{eq:25}
    \partial_t\vartheta + ({\bm v}\nabla)\vartheta
    =     \kappa\Delta \vartheta + f,
\end{equation}
where $\bm v$ is the flow velocity, $f$ is the external source of the scalar, $\kappa$ is its molecular diffusion coefficient, and $\Delta$ designates Laplacian. We assume that the flow velocity $\bm v$ has a random component that is small compared to its constant part and  forcing $f$ is a stochastic quantity which has characteristic scale $L$ in space.  We assume that the influence of the diffusion at scale $L$ is weaker not only than one from the constant part of the velocity, but also than the effect of stirring acceleration caused by the random part of the velocity. All the criteria are formulated below, see~(\ref{DllSigma},\ref{large-Peclet-effective}).

For our limit of weak diffusion, it is reasonable to study first separately the evolution of the passive scalar in the absence of it. If one neglects the diffusion term in Eq. (\ref{eq:25}), then its solution can be written in terms of Lagrangian trajectories $\bm q (t)$ that are governed by the equation
\begin{equation}
\partial_t \bm q (t) = \bm v(t,\bm q).
\label{lagrtraj}
\end{equation}
The solution of the diffusionless equation (\ref{eq:25}) with the initial condition $\vartheta(0,\bm r)$, taken at $t=0$, is
\begin{eqnarray}
\vartheta(t, \bm r) =
\vartheta[0, \bm q(0)]
+\int_0^t dt_1\, f[t_1, \bm q(t_1)].
\label{solula1}
\end{eqnarray}
Here $\bm q$ is the Lagrangian trajectory passing through the point $\bm r$ at the time $t$, ${\bm q}(t)={\bm r}$.

Further in the work we examine the case where the velocity field $\bm v$ is smooth that is it can be expanded into Taylor series with the convergence radius larger than all scales characterizing the passive scalar evolution. The scalar spatial distribution is influenced mainly by the smooth component of the flow, whereas the effect of its relatively weak small-scale fluctuations can be included into renormalization of the diffusion coefficient $\kappa$~\cite{shraiman1994lagrangian}.

\subsection{Statistics of Lagrangian trajectories}
\label{subsec:lagrange}

Let us examine statistical properties of the difference $\bm\ell=\bm q_1-\bm q_2$ between two Lagrangian trajectories. Our interest is the probability density function (PDF) for $\bm \ell$ at different times assuming some fixed initial value of $\bm\ell$ or its initial probability distribution. For scalar as ensemble of blobs with homogeneous spatial statistics the PDF can be thought as probability for both starting and ending points of $\bm \ell$ getting inside of it. Vectors $\bm \ell$ and $-\bm \ell$ are equivalent for the description of the scalar spatial distribution in this sense, thus $\bm \ell$ can be called director. First we formulate the dynamical equation for $\bm\ell$, and then extract its statistical properties by averaging over the statistics of the random flow.

Assuming that the difference $\bm\ell$ lies inside the region of the smoothness of the velocity field, we find from Eq. (\ref{lagrtraj})
\begin{equation}
\partial_t \ell_i  = \ell_k \partial_k v_i,
\label{lagrtraj2}
\end{equation}
where we kept the main term of the expansion of the velocity field $\bm v$ in Taylor series. The velocity gradient $\partial_k v_i$ in Eq. (\ref{lagrtraj2}) is a function of time, determined by the structure of the velocity field in the vicinity of the Lagrangian trajectories $\bm q_1$ and $\bm q_2$,  $\bm\ell=\bm q_1-\bm q_2$.

Further we focus on the two-dimensional dynamics. We examine the case where the velocity field $\bm v$ contains both the regular (deterministic) contribution and the fluctuating (random) one. The regular contribution is assumed to be a shear flow. We chose the axes $X,Y$ of the reference frame to fix the shear flow velocity as $v_x=\Sigma y$, where $\Sigma$ is the shear rate which is presumed to be positive for definiteness. Then we obtain from Eq.~(\ref{lagrtraj2})
\begin{eqnarray}
\label{elldynamics1}
\partial_t \ell_x =\Sigma \ell_y+\ell_x\partial_xu_x+ \ell_y\partial_yu_x
 \\ \label{elldynamics2}
\partial_t \ell_y=\ell_x\partial_xu_y+ \ell_y\partial_yu_y,
\end{eqnarray}
where $\bm u$ is the fluctuating part of the velocity. Its statistical properties are assumed to be homogeneous in time and space.

The fluctuating part of the velocity $\bm u$ is supposed to be relatively weak. To characterize the weakness, one introduces the tensor
\begin{equation}
    D_{ijkl}=
    \int_0^\infty dt\, \langle \partial_iu_j(t) \partial_ku_l(0) \rangle,
\label{angulard}
\end{equation}
All elements of the tensor $D$ are assumed to be of the same order. The angular brackets in Eq. (\ref{angulard}) and subsequently denote averaging over statistics of the random flow $\bm u$. As it will be shown below, the only element of our interest is $D=D_{xyxy}$ due to the anisotropy dictated by the shear flow. The weakness of the random flow in comparison with the shear flow means
\begin{equation}\label{DllSigma}
    D\ll \Sigma.
\end{equation}

The dynamics of the vector $\bm \ell$ is peculiar~\cite{chertkov2005polymers} due to fluctuations of $\ell_y$, caused by the random part of the velocity, see Eq. (\ref{elldynamics2}). The main term in right hand side of Eq. (\ref{elldynamics1}) is $\Sigma \ell_y$. Therefore, if $\ell_y>0$ then $\ell_x$ grows towards positive values. However, if $\ell_y$ becomes negative, $\ell_x$ starts to diminish and then changes its sign and grows towards negative values. This process when $\ell_x$ changes abruptly its sign is called tumbling. Precise streamline alignment of $\bm{\ell}$ is an unstable stationary point for the vector direction in dynamics with constant shear only. The fluctuating component of the flow enables tumbling processes by changing $\ell_y$ sign. Tumblings occur aperiodically in a characteristic time $D^{-1/3}\Sigma^{-2/3}$.

Between the tumblings $\ell_x \gg \ell_y$. Ratio $\ell_y/\ell_x$ can be estimated as $(D/\Sigma)^{1/3}\ll1$ then, being the characteristic angle between the director and the streamlines. However, during the each tumbling $\ell_x$ diminishes by a large factor. To avoid particular analysis of the tumbling processes, we exploit the following parametrization of the vector $\bm\ell$
\begin{equation}
(D/\Sigma)^{1/3}\ell_x= l_0 \exp\varrho \cos\phi, \quad
\ell_y=l_0 \exp\varrho \sin\phi,
\label{homog5}
\end{equation}
where $l_0$ is a constant determined by the initial value of $\bm\ell$. Then the inequality $(D/\Sigma)^{1/3} \ell_x \lesssim \ell_y$ is satisfied for most of the time. Therefore, the quantity $\varrho$ does not experience strong changes unlike~$\ell_x$. Angle $\phi+\pi$ corresponds to inversion of vector $\bm\ell$ and thus is equivalent to $\phi$, so it will be enough to consider angle on the interval $-\pi/2<\phi<\pi/2$, treating functions of $\phi$ (say, PDF of $\phi$) as periodic with a period $\pi$.

Substituting the expressions (\ref{homog5}) into Eqs. (\ref{elldynamics1},\ref{elldynamics2}) one concludes that due to the inequality $D\ll\Sigma$ (\ref{DllSigma}) the only relevant component of the gradients of the random velocity is $\partial_x u_y$. Thus, we come to the stochastic system:
\begin{eqnarray}
\partial_\tau \varrho= (1+\zeta) \cos\phi \sin\phi,
\label{homog6}\\
\partial_\tau \phi= \zeta \cos^2 \phi - \sin^2\phi,
\label{homog7}
\end{eqnarray}
where $\zeta(\tau)=\Sigma^{-1/3}D^{-2/3} \partial_x u_y$ and we have introduced the dimensionless time
\begin{equation}
\tau=\Sigma^{2/3}D^{1/3}t.
\label{taudef}
\end{equation}
Further, we present all relations in terms of the  dimensionless time $\tau$.

In the equations (\ref{homog6},\ref{homog7}), the regular and the random terms in right hand sides are comparable, which was our motivation to introduce the parametrization (\ref{homog5}). Note that Eq. (\ref{homog7}) is a closed stochastic equation for the angle $\phi$, that is a consequence of linearity of the equations (\ref{elldynamics1},\ref{elldynamics2}). Thus, one can independently examine statistical characteristics of the angle $\phi$, based on Eq. (\ref{homog7}). The variable $\varrho$ grows in average as time goes. The growth can be characterized by the dimensionless Lyapunov exponent $\lambda =\langle \partial_\tau \varrho \rangle$ (the dimensional Lyapunov exponent is $\lambdabar =\lambda (D\Sigma^2)^{1/3}$).  One can also introduce the quantity $\omega=-\langle \partial_\tau \phi \rangle$, that is the dimensionless frequency of the tumbling processes, so tumblings occur an average of $\pi (\Sigma D^2)^{1/3}t/\omega$ times in the system for a large $t$. Both quantities, $\lambda $ and $\omega$, are of order of unity.

The general structure of the stochastic equations (\ref{homog6},\ref{homog7}) enables one to find a relation for $\Pi$ if the statistics of $\zeta$ is invariant under the time invertion. The property is assumed below. Then one relates the values of $\Pi(\varrho)$ for different signs of $\varrho$:
\begin{equation}
\Pi(-\varrho)=\exp(-2\varrho) \Pi(\varrho).
\label{symme}
\end{equation}
The relation (\ref{symme}) implies that at $\tau=0$ the angle $\phi$ is fixed and that $\varrho=0$. The proof of the relation (\ref{symme}) can be found in Appendix \ref{sec:symmetry}.

Asymptotically, at $\tau\to\infty$, PDF $P(\phi)$ turns to a stationary distribution, if $\zeta$ has statistical properties homogeneous in time. As for PDF $\Pi(\varrho)$, it does not turn stationary at large times $\tau$ since $\varrho$ grows in average. Instead, in accordance with the theory of large deviations \cite{feller1968probability,klenke2007probability} at $\tau\gg1$
\begin{equation}
\Pi(\varrho)
\propto \exp\left[-\tau S(\varrho/\tau)\right],
\label{kramers}
\end{equation}
where $S(\xi)$ is the so-called Cram\'{e}r (or entropy) function, which is convex. The function has minimum at $\xi=\lambda $, that is $\varrho=\lambda  \tau = \lambdabar t$. Hence
\begin{equation}
S'(\lambda )=0,
\label{minimum}
\end{equation}
where $S'\equiv dS/d\xi$. The normalization in Eq. (\ref{kramers}) is determined by a close vicinity of the minimum point. We assume that $S(\lambda )=0$. Then the normalized function
\begin{equation}
    \Pi(\varrho)
    =\sqrt{\frac{S''(\lambda )}{2\pi \tau}}
\exp\left[-\tau S(\varrho/\tau)\right]
\label{weightcr}
\end{equation}
is valid at $\tau\gg1$.

The general law (\ref{symme}) leads to the relation
\begin{equation}
S(-\xi)=S(\xi)+2\xi,
\label{kramers3}
\end{equation}
as it follows from Eq. (\ref{kramers}). Taking the derivative of the relation (\ref{kramers3}) one obtains
\begin{equation}
S'(-\xi)= -S'(\xi)-2.
\label{cramers3}
\end{equation}
Substituting here $\xi=0$, one obtains
\begin{equation}
S'(0)=-1.
\label{homog16}
\end{equation}
Another consequence of Eq. (\ref{cramers3}) is $S'(-\lambda )=-2$, that can be established using Eq. (\ref{minimum}).

It is instructive to introduce Fourier transform of $\Pi(\varrho)$,
\begin{equation}
\widetilde \Pi (\eta)=\int d\varrho\, \exp(-\eta \varrho) \Pi(\varrho).
\label{fourier1}
\end{equation}
In the conventional Fourier transform $\eta$ is purely imaginary. However, we treat $\eta$ as an arbitrary complex number. In the limit $\tau\gg1$ we can use the expression (\ref{kramers}) and the integral (\ref{fourier1}) can be taken in the saddle point approximation. As a result, we find
\begin{equation}
\widetilde \Pi (\eta ) \propto \exp[-\gamma(\eta )\tau]
\label{fourier2}
\end{equation}
where the function $\gamma(\eta )$ is related to the Cram\'{e}r function $S(\xi)$ via the Legendre transform
\begin{eqnarray}
S=\gamma-\eta\xi,
\label{homog14} \\
\partial_\eta  \gamma =\xi, \quad
\partial_\xi S=-\eta .
\label{homog15}
\end{eqnarray}
Solutions of Eqs. (\ref{homog14},\ref{homog15}) correspond to real $\eta$. Taking into account the relation (\ref{cramers3}) one concludes that (\ref{kramers3}) is equivalent to
\begin{equation}
\gamma(\eta )=\gamma(2-\eta).
\label{twoxi}
\end{equation}

The dimensionless Lyapunov exponent $\lambda $ is equal to the ratio $\xi=\varrho/\tau$, taken at the minimum of the Cram\'{e}r function $S(\xi)$. As it follows from Eq. (\ref{homog15}), the minimum of $S$ is achieved at $\eta =0$. Thus, we find from Eq. (\ref{homog15})
\begin{equation}
\lambda  =\partial_\eta  \gamma(0).
\label{lambda2}
\end{equation}
Since $\gamma$ is invariant under the transformation $\eta \to 2-\eta $, the derivative of $\gamma$ over $\eta $ at $\eta =1$ is equal to zero, $\partial_\eta  \gamma(1)=0$. Thus, we conclude from Eq. (\ref{homog14}), that point $\eta =1$ ($\varrho=\lambda  \tau$) corresponds to the value $\xi=0$.

\subsection{Random flow short correlated in time}

To demonstrate main features of the statistics of the Lagrangian trajectories, we examine the model where the random flow is short correlated in time. The model enables one to draw a number of analytical results \cite{turitsyn2007polymer}. In terms of our parametrization (\ref{homog5}), the model is determined by the pair correlation function
\begin{equation}
\left\langle \zeta(\tau_1) \zeta(\tau_2) \right\rangle
=2 \delta(\tau_1-\tau_2).
\label{homog4}
\end{equation}
The factor in the right hand side of Eq. (\ref{homog4}) is written in accordance with Eq. (\ref{angulard}).

We find, as a consequence of Eqs. (\ref{homog6},\ref{homog7}) that the dimensionless Lyapunov exponent and the dimensionless frequency of the tumblings are equal to
\begin{eqnarray}
\lambda =\langle \partial_\tau \varrho \rangle=
\langle \cos\phi\sin\phi+\cos(2\phi)\cos^2\phi \rangle,
\label{lambda} \\
\omega=-\langle \partial_\tau \phi \rangle
=\langle \sin^2\phi+\sin(2\phi) \cos^2\phi \rangle.
\label{tumbling}
\end{eqnarray}
The first terms in the angular brackets in Eqs. (\ref{lambda},\ref{tumbling}) are related to the regular terms in  right hand sides of Eqs. (\ref{homog6},\ref{homog7}), whereas the second terms in the angular brackets are related to the terms with random variable $\zeta$ there. To find the latter contributions, one should find increments of $\varrho,\phi$, caused by $\zeta$, take into account the increments in the right hand sides of Eqs. (\ref{homog6},\ref{homog7}) and then average the products of the increments and $\zeta$, using Eq. (\ref{homog4}). The quantities (\ref{lambda},\ref{tumbling}) are expressed in terms of the statistics of the angle $\phi$ and can be calculated irrespective to the statistics of $\varrho$.

The Langevin equations (\ref{homog6},\ref{homog7}) with the random variable governed by Eq. (\ref{homog4}) enable one to establish the Fokker-Planck equations either for the PDF of the variable $\phi$ only or for the joint PDF of the variables $\varrho,\phi$. The corresponding technique is well known \cite{oksendal2003stochastic,klebaner2005introduction}. Therefore, we do not present the derivation of the  Fokker-Planck equations, focusing on analysing their solutions.

\begin{figure}[t]
\includegraphics[width=\linewidth]{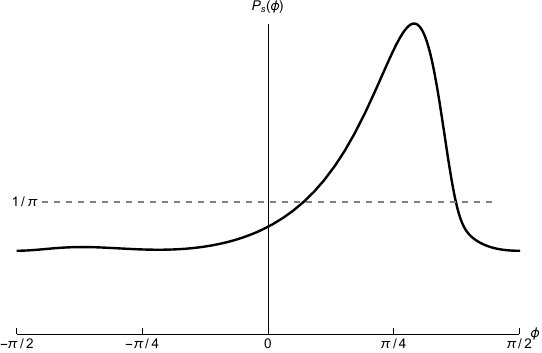}

\caption{Stationary PDF $P_s$ of the angle $\phi$ in the case of the short correlated in time flow fluctuations (\ref{homog4}).}
\label{fig:P_phi_stat}
\end{figure}

We begin with the Fokker-Planck equation for PDF of $\phi$, $P(\phi)$. The equation follows from Eqs. (\ref{homog7},\ref{homog4}):
\begin{eqnarray}
\partial_\tau{ P}=\partial_\phi \left( \sin^2\phi\, { P}\right)
+\partial_\phi \left[ \cos^2 \phi\, \partial_\phi \left( \cos^2 \phi \, { P}\right) \right].
\label{homog8}
\end{eqnarray}
The equation (\ref{homog8}) has to be supplemented by the periodicity condition in terms of the angle $\phi$ and by some initial condition. Say, the initial angle has some fixed value $\phi_0$, that leads to the initial $\delta$-function, $P=\delta(\phi-\phi_0)$ (continued periodically with the period $\pi$).

At $\tau\to \infty$ a stationary PDF of $\phi$ is achieved. The stationary solution $P_s$ of the equation (\ref{homog8}) is written as
\begin{equation}
P_s(\phi)=\frac{N}{\cos^{2}\phi}\int\limits_0^{\pi/2}\frac{d\psi}{\cos^{2}\psi}
e^{\left[\left(\tan\phi-\tan\psi\right)^{3}-\tan^{3}\phi\right]/3},
\label{P_phi_stat}
\end{equation}
where the constant \hbox{$N=3^{5/6}/\left[2^{1/3}\sqrt{\pi}\Gamma\left(1/6\right)\right]$} is found from the normalization condition $\int d\phi\, P_s=1$. The function $P_s$ is plotted in Fig.~\ref{fig:P_phi_stat}. Calculating numerically the averages (\ref{lambda},\ref{tumbling}) as integrals over $\phi$ with the weight $P_s(\phi)$, one obtains
\begin{equation}
    \lambda     =
    \frac{\pi N}{\sqrt{3}} \approx 0.36 ,
    \qquad
    \omega=\pi N \approx 0.63.
\label{lamom}
\end{equation}
The numerical values (\ref{lamom}) are in agreement with the analysis given in Ref.~\onlinecite{turitsyn2007polymer}, see also Appendix~\ref{app:rho-psi} for more detailed comparison of our mathematical approach with the one
used there.

Next, we turn to the joint PDF for $\varrho,\phi$, ${\mathcal P}(\tau, \varrho,\phi)$. The Fokker-Planck equation for the quantity is
\begin{eqnarray}
\partial_\tau{\mathcal P}
=\partial_\phi \left( \sin^2\phi\, {\mathcal P}\right)
-\partial_\varrho \left( \cos\phi\sin\phi\, {\mathcal P}\right)
\nonumber \\
+\partial_\phi \left[ \cos^2 \phi\, \partial_\phi \left( \cos^2 \phi \, {\mathcal P}\right) \right]
+\cos^2\phi \sin^2\phi\, \partial_\varrho^2 {\mathcal P}
\nonumber \\
+ \partial_\phi \left( \cos^3\phi \sin\phi\, \partial_\varrho {\mathcal P} \right)
+\cos\phi \sin\phi \partial_\phi \left( \cos^2\phi\, \partial_\varrho {\mathcal P}\right),
\label{homog9}
\end{eqnarray}
as a consequence of Eqs. (\ref{homog6},\ref{homog7},\ref{homog4}). PDF of $\varrho$, $\Pi(\varrho)$, can be found as
\begin{equation}
\Pi(\varrho)=\int_{-\pi}^{\pi}\frac{d\phi}{2\pi}
{\mathcal P}(\varrho,\phi).
\nonumber
\end{equation}
Note that no closed equation for $\Pi$ can be derived from Eq. (\ref{homog9}).

We see that the equation (\ref{homog9}) is homogeneous in $\varrho$, it is a direct consequence of the linearity of the initial equations (\ref{elldynamics1},\ref{elldynamics2}). Therefore it is worth to analyze the equation in terms of Fourier transform of ${\mathcal P}(\varrho)$. We introduce it by analogy with Eq. (\ref{fourier1}):
\begin{equation}
\widetilde{\mathcal P}=\int d\varrho\, \exp\left(-\eta \varrho\right) {\mathcal P},
\label{homog10}
\end{equation}
where $\eta$ is some (complex) parameter. The analysis of the object $\widetilde {\mathcal P}$ can be found in Appendix \ref{sec:fourier}. As a result, one can extract the function $\gamma(\eta )$ introduced by Eq. (\ref{fourier2}) and check directly the property (\ref{twoxi}).

The function $\gamma(\eta)$ can be calculated numerically. One can check that the value of $\lambda $, calculated in accordance with Eq. (\ref{lambda2}) coincides with one given by Eq. (\ref{lamom}). Converting $\gamma(\eta)$ into $S(\xi)$ in accordance with Eqs. (\ref{homog14},\ref{homog15}), one finds that in the minimum of $S(\xi)$, where $\xi=\lambda $, \begin{equation}
    S''\approx 2.46 , \quad
    S'''\approx-1.56 .
\label{homog17}
\end{equation}
The values (\ref{homog17}) enables one to approximate the Cram\'{e}r function $S$ near its minimum and to find the factor in Eq. (\ref{weightcr}).

\section{Decay of the scalar}
\label{sec:decay}

Here we consider decay of the passive scalar, which is described by the basic equation (\ref{eq:25}) with $f=0$. We are interested in evolution of correlation functions of the passive scalar $\vartheta$ that have to be obtained by averaging over the statistics of the random flow. The statistical properties of $\vartheta$ appear to be extremely non-Gaussian at $\tau\gg1$, where the dimensionless time $\tau$ is introduced by Eq. (\ref{taudef}). We establish some features of the statistics.

Further we examine quantities obtained by averaging over an ensemble of the realizations of the initial distributions $\vartheta(0,{\bm r})$. We consider each one as the aggregation of similar blobs of the scalar fluctuations placed in the flow at $t=0$, keeping zero total amount of scalar. Assuming the limit of their high concentration, i.e. overlapping of many blobs in each point, value of $\vartheta(0,\bm{r})$ is a sum of large number of independent variables. Thus, as a consequence of the central limit theorem, the field $\vartheta(0,\bm{r})$ possesses Gaussian statistics with zero mean~\cite{vergeles2006spatial}.

Let us introduce the object ${\mathcal F}(t,{\bm r}_1,\bm r_2)$ that is the product $\vartheta(t,\bm r_1) \vartheta(t,\bm r_2)$ averaged over the statistics of the initial values of $\vartheta$. To find any correlation function of the scalar, one should take the product $\vartheta(\bm r_1) \vartheta(\bm r_2) \dots$ and average it first over the initial statistics and then over the statistics of the random velocity field. The first step reduces the product $\vartheta(\bm r_1) \vartheta(\bm r_2) \dots$ to the product of ${\mathcal F}$'s with some combinatoric factor in accordance with Wick theorem \cite{wick1950theorem}. To make the second step, one should establish statistical properties of ${\mathcal F}$. We proceed to the problem.

If the ensemble of the initial values is statistically homogeneous in space, then ${\mathcal F}(t,{\bm r}_1,\bm r_2)$ is a function solely of the difference $\bm r=\bm r_1-\bm r_2$, ${\mathcal F} = {\mathcal F}(t,\bm r)$. In the case, one finds from Eq.  (\ref{eq:25})
\begin{equation}
\partial_t {\mathcal F} +\Sigma y \partial_x {\mathcal F}
+(\partial_\beta u_\alpha) r_\beta \partial_\alpha {\mathcal F}
=2\kappa \nabla^2{\mathcal F},
\label{pairde}
\end{equation}
where we have presumed as above the flow is smooth and consist of the shear flow with the velocity $v_x=\Sigma y$ and the random flow with the velocity $\bm u$. Let us stress that the object ${\mathcal F}$ is a functional of the random variable $\partial_\beta u_\alpha$, entering the equation (\ref{pairde}).

By analogy with the analysis of the Lagrangian trajectories, see Section \ref{subsec:lagrange}, we pass to the rescaled coordinate $w=(D/\Sigma)^{1/3}x$ and the dimensionless time (\ref{taudef}). Then one finds from Eq. (\ref{pairde})
\begin{equation}
\partial_\tau {\mathcal F}+y \partial_w {\mathcal F}
+\zeta w \partial_y {\mathcal F}
=r_\kappa^2 \partial_y^2{\mathcal F},
\label{thetadim}
\end{equation}
where $\zeta=\Sigma^{-1/3}D^{-2/3} \partial_x u_y$. We have kept in Eq. (\ref{thetadim}) the only relevant component of the random velocity gradient, $\partial_x u_y$, the main derivative $\partial_y$ in Laplacian, and have introduced the diffusive scale
\begin{equation}
    r_\kappa
    =
    (2\kappa)^{1/2}\Sigma^{-1/3}D^{-1/6}.
\label{rkappa}
\end{equation}
Batchelor scale $r_\kappa$ is assumed to be much smaller than characteristic scales of the initial scalar field and of the forcing one.

It is instructive to examine a Gaussian shape of ${\mathcal F}$: such profile of spatial distribution, formed by the initial statistics, is preserved in Eq. (\ref{pairde}). We suppose that initially ${\mathcal F}(0,\bm r)\propto \exp(-r^2/L^2)$, where $L$ is the characteristic initial scale. Then the quantity ${\mathcal F}$ at any time $t$ is expressed as
\begin{equation}
\label{eq:02}
    {\mathcal F}
    =
    \frac{\sqrt{\det \hat\Lambda}}{\sqrt{\det \hat\Lambda\big\vert_{t=0}}}
    \exp\left(-\Lambda_{\alpha\beta}b_\alpha b_\beta\right),
\end{equation}
where the $\bm b$ is a coordinate vector in rescaled space, $b_\alpha=(w,y)$, the symmetric matrix $\hat \Lambda$ is a function of time, which dynamics is consistent with Eq. (\ref{thetadim}), and we have normalized the scalar intensity so ${\mathcal F}(0,0)=1$. The time-dependent factor at the exponent in Eq. (\ref{eq:02}) is determined by the fact that total amount of the passive scalar $\int\mathrm{d}^2 b \, \vartheta$ is conserved in time according to Eq. (\ref{thetadim}), see Ref.~\onlinecite{corrsin1951decay}.

We use the following parametrization of the matrix $\hat \Lambda$ figuring in Eq. (\ref{eq:02})
\begin{eqnarray}
\hat\Lambda=
\left(\begin{array}{cc}
c & -s \\
s & c
\end{array}\right)
\left(\begin{array}{cc}
L_+^{-2} & 0 \\
0 & L_-^{-2}
\end{array}\right)
\left(\begin{array}{cc}
c & s \\
-s & c
\end{array}\right),
\label{matrixp}
\end{eqnarray}
where $c=\cos \phi$, $s=\sin \phi$. The eigenvalues of the matrix (\ref{matrixp}) are $L_+^{-2}$, $L_-^{-2}$, where $L_{\pm}$ can be interpreted as sizes of the scalar blob in ${\bm b}$-space. Therefore the factor $\sqrt{\det \hat\Lambda}$ entering Eq. (\ref{eq:02}) is equal to
\begin{equation}
\sqrt{\det \hat\Lambda}= (L_+ L_-)^{-1},
\label{factor}
\end{equation}
i.e. inverse area occupied in ${\bm b}$-space by the blob. In accordance with Eqs. (\ref{eq:02},\ref{matrixp}), for the initial profile $\propto \exp(-r^2/L^2)$, we have $\phi(0)=\pi/2$,
\begin{equation}
L_+(0)=L, \quad L_-(0)=L_\star, \quad
L_\star=(D/\Sigma)^{1/3}L.
\label{initialv}
\end{equation}
Thus, initially $L_+\gg L_-$. We will see that the ratio $L_+/L_-$ typically grows with time and will neglect very rare events when the ratio becomes of order unity. Thus, we assume $L_+\gg L_-$ is fulfilled further all the time.

Substituting the parametrization (\ref{matrixp}) into Eq. (\ref{eq:02}) and then using Eq. (\ref{thetadim}), one finds the equations for the angle $\phi$ and the parameters $L_\pm$. In the limit $L_+\gg L_-$ one reproduces the equation (\ref{homog7}) for the angle $\phi$ and the equations for $L_\pm$ are
\begin{eqnarray}
\partial_\tau \ln L_+=\cos\phi\, \sin\phi\, (1+\zeta),
\label{lplus} \\
\partial_\tau \ln L_-=-\cos\phi\, \sin\phi\, (1+\zeta)
+2 \frac{r_\kappa^2}{L_-^{2}}\cos^2\phi.
\label{lminus}
\end{eqnarray}
Our interest is statistics of the solutions of Eqs. (\ref{lplus},\ref{lminus}) at times $\tau\gg1$ where PDF of the angle $\phi$ achieves stationary distribution.

The equation (\ref{lplus}) coincides with Eq. (\ref{homog6}) for $\varrho$. Consequently, $\ln (L_+/L)$ has the same statistical properties as $\varrho$, see Section \ref{sec:general}. Typically, $\ln (L_+/L)$ is estimated as $\lambda  \tau$. As to the quantity $L_-$, its statistical properties depend on its value. If $L_-\gg r_\kappa$ then the last term in Eq. (\ref{lminus}) is irrelevant, and we find $L_- =  L L_\star/L_+$. Therefore $L_-$ typically diminishes exponentially as time goes. If $L_-$ reaches $r_\kappa$, then its statistical properties become stationary, and the estimate $L_-\sim r_\kappa$ is valid.

The duration of the first (advective or diffusionless) stage is
\begin{equation}
\tau_\kappa =\frac{1}{2\lambda }\ln \frac{D L^2}{\kappa}.
\label{duration}
\end{equation}
We assumed here
\begin{equation}\label{large-Peclet-effective}
    D L^2/\kappa \gg 1.
\end{equation}
More precisely, below we assume that $\ln(D L^2/\kappa)$ is large, so $\tau_\kappa \gg1$. Otherwise, if $DL^2/\kappa \lesssim 1$, the mixing process is determined only by the mean shear flow $\Sigma$ and the molecular diffusion, the limit $\Sigma L^2/\kappa\gg1$ was considered in \cite{souzy2018mixing}. Inequality (\ref{large-Peclet-effective}) implies that at scale $L$ the influence of the molecular diffusion is weaker than both the influence of shear flow and the stirring acceleration by the flow's random component. Note that limit of the large P\'{e}clet number $\mathrm{Pe}=L^2/r_\kappa^2\sim (D\Sigma^2)^{1/3}L^2/\kappa\gg 1$ in our system does not provide a sufficient criterion. The amplitude (\ref{factor}) remains constant at the advective stage and behaves $\propto L_+^{-1}$ at the second (diffusive) stage. Since initially $L_+L_-= L L_\star$ we conclude that $L_+\sim L L_\star/ r_\kappa$ in the transition region between the stages, which leads to the condition
\begin{equation}
\ln(L_+/L)>\lambda  \tau_\kappa
\label{varrhose}
\end{equation}
at the diffusive stage.

\subsection{Single point statistics}
\label{subsec:one-point-mean}

Let us analyze moments of the passive scalar that are single-point means $\langle |\vartheta|^{2\alpha} \rangle$. To find the moments we use the expression
\begin{equation}
\langle |\vartheta|^{2\alpha} \rangle=C_\alpha \langle [{\mathcal F}(t,\bm 0)]^\alpha \rangle,
\label{moments}
\end{equation}
where $C_\alpha = 2^{\alpha}\Gamma(\alpha+1/2)/\sqrt{\pi}$. The expression (\ref{moments}) is a consequence of the initial Gaussian statistics of $\vartheta(0,\bm{r})$. Although the expression (\ref{moments}) implies a particular statistics of initial values of $\vartheta$, results concerning the behavior of the moments at large times $\tau \gg 1$ are universal, because it is determined by the statistics of the flow fluctuations.

Substituting the expression (\ref{eq:02}) with the factor (\ref{factor}) into Eq. (\ref{moments}) one arrives at the following expression for the moments
\begin{equation}
\label{eq:07}
    \langle |\vartheta|^{2\alpha}\rangle
    =
    C_\alpha \left \langle (L_+L_-/L L_\star)^{-\alpha}\right\rangle.
\end{equation}
Thus, the passive scalar moments can be calculated using the statistical properties of $L_\pm$ established above. At the advective stage, where $\tau<\tau_\kappa$, the product $L_+L_-$ remains constant and mean $\langle |\vartheta|^{2\alpha}\rangle$ is independent of time $\tau$. Further we focus on the diffusive stage where diffusion is relevant. Then $L_-$ is estimated as $r_\kappa$ whereas the typical value of $L_+$ depends on $\alpha$ and $\tau$.

Since the factor (\ref{factor}) is proportional to $L_+^{-1}$  at the diffusive stage, one can write
\begin{equation}\label{seconds}
    \langle |\vartheta|^{2\alpha}\rangle
    \sim
    \int d\varrho \exp\left[-\alpha (\varrho -\lambda  \tau_\kappa)
    -\tau S(\varrho/\tau)\right],
\end{equation}
where $\varrho=\ln(L_+/L)$ and we have exploited Eq. (\ref{weightcr}), omitting its pre-exponential factor as well as $C_\alpha$ in (\ref{eq:07}) as a multiplier with weak dependence on $\alpha$. In (\ref{seconds}), we have subtracted from $\varrho$ its value in the transition region between the stages, see Eq. (\ref{varrhose}), since the moment is unchanged at the advective stage.

In case when $0<\alpha<1$ at large $\tau$ the integral (\ref{seconds}) for moment $\langle |\vartheta|^{2\alpha}\rangle$ is determined by the saddle point - solution of the equation
\begin{equation}
\alpha+ S'(\xi)=0.
\label{second2}
\end{equation}

For $0<\alpha<1$ the value of $\xi$, found in accordance with Eq. (\ref{second2}), lies in the interval $0<\xi<\lambda $, since $S'(\lambda )=0$ and $S'(0)=-1$, see Section \ref{subsec:lagrange}. Hence, the result is determined with $\varrho=\xi \tau$ and by (\ref{varrhose}), that is $\varrho>\lambda \tau_\kappa$, so time $\tau$ should satisfy the inequality $\tau>\lambda  \tau_\kappa /\xi$:
\begin{equation}\label{second3}
    \langle |\vartheta|^{2\alpha}\rangle
    \sim
    \exp[\alpha \lambda  \tau_\kappa-\gamma(\alpha) \tau],
\end{equation}
where $\gamma(\eta )$ is found in accordance with the Legendre transform (\ref{homog14},\ref{homog15}). For smaller $\tau$ the integral (\ref{seconds}) is determined by the smallest possible value $\varrho=\lambda  \tau_\kappa$. Then we find
\begin{equation}\label{second4}
    \langle |\vartheta|^{2\alpha}\rangle
    \sim
    \exp\left[-\tau S(\lambda  \tau_\kappa/\tau) \right].
\end{equation}
The expression is valid if $\tau_\kappa<\tau< \lambda  \tau_\kappa /\xi$. Note that (\ref{second4}) is independent of $\alpha$ since it gives the probability that $\varrho$ does not grow as time goes.

If $\alpha>1$, the condition $\varrho>\lambda \tau_\kappa$ is violated for the saddle point, determined by Eq. (\ref{second2}), for any $\tau$. Indeed, the equation leads to $\xi<0$ and, consequently, to $\varrho<0$. In this situation, again, the integral (\ref{seconds}) is determined by the smallest possible value $\varrho=\lambda  \tau_\kappa$. We conclude, that the relation (\ref{second4}) is correct at any time $\tau>\tau_\kappa$ for $\alpha>1$.

\subsection{Correlation functions}
\label{subsec:pair}

The pair correlation function $F$ of the passive scalar $\vartheta$ can be written as the average
\begin{equation}
F(t,\bm r)=\langle {\mathcal F}(t,{\bm r}) \rangle.
\label{pair1}
\end{equation}
We suppose that the initial statistics leads to the Gaussian initial form of ${\mathcal F}$, so it has form as in Eq. (\ref{eq:02}) at any time. The parametrization  (\ref{matrixp}) implies that averaging in Eq. (\ref{pair1}) is performed over the statistics of $\phi, L_+,L_-$, examined above.

Passing to polar coordinates in the rescaled space:
\begin{equation}
w=(D/\Sigma)^{1/3} x=b \cos\psi, \quad
y=b\sin\psi,
\label{pair2}
\end{equation}
we find for the argument in the exponent in Eq. (\ref{eq:02})
\begin{equation}
\Lambda_{\alpha\beta}b_\alpha b_\beta
=b^2 L_+^{-2}\cos^2(\phi-\psi)
+b^2 L_-^{-2} \sin^2(\phi-\psi).
\label{pair3}
\end{equation}
Thus, if $b\ll L_-$ then the quantity (\ref{pair3}) is much less than unity and the pair correlation function is reduced to the second moment of $\vartheta$ examined in Section \ref{subsec:one-point-mean}. Hence below we examine the case $b \gg L_-$. Note that the criterion depends on time at the advective stage of the passive scalar evolution and is reduced to $b\gg r_\kappa$ at the diffusive stage.

If $b \gg L_-$ then the quantity (\ref{pair3}) has the deep minimum at $\phi=\psi$ and, correspondingly, $\exp(-\Lambda_{\alpha\beta}b_\alpha b_\beta)$ has the sharp peak at this point. Thus, averaging over $\phi$ statistics brings to the factor proportional to the peak's width,
\begin{equation}\label{pair4}
    F\sim \left\langle \frac{1}{bL_+}\exp(-b^2 L_+^{-2}) \right\rangle_+,
\end{equation}
as a consequence of Eqs. (\ref{eq:02},\ref{factor}). Remarkably, $L_-$ falls out of the consideration and we stay with averaging solely over the statistics of $L_+$ in Eq. (\ref{pair4}). The factor in the proportionality law (\ref{pair4}) depends on the angle $\psi$. Its exact value is determined by details of the $\phi$ distribution and, consequently, is not universal, since that depends on the statistics of $\zeta$ (\ref{homog7}). As we have implemented the rescaling (\ref{homog5},\ref{pair2}), the distribution is anticipated to be nearly isotropic, an example for short correlated case is given in Fig.~\ref{fig:P_phi_stat}. For this reason we focus on the dependence of the pair correlation function on the length $b$.

The average in (\ref{pair4}) can be written as the integral over $\varrho=\ln(L_+/L)$ with the weight (\ref{weightcr}). Due to the strong dependence on $\varrho$ of $\exp(-b^2 L_+^{-2})$, the last factor restricts the integration region to $\varrho>\ln (b/L)$, so we arrive to
\begin{equation}\label{pair5}
    F
    \sim
    \frac{L_\star}{b}\int_{\ln(b/L)}^\infty d\varrho
    \exp\left[-\varrho -\tau S(\varrho/\tau)\right]
\end{equation}
if $b\gg L_-$. The integral above is of the same type as in Section \ref{subsec:one-point-mean} and can be analyzed similarly.

If $\ln(b/L)<0$ ($b\ll L$) then the integral in Eq. (\ref{pair5}) is determined by the saddle point $\varrho=0$ in accordance with (\ref{homog16}) and we obtain
\begin{equation}
    F\sim \frac{L_\star}{b}
\exp\left[ -\tau S(0)\right].
\label{pair6}
\end{equation}
If $\ln(b/L)>0$ ($b\gg L$) then the integral in Eq. (\ref{pair5}) is determined by the minimal value of $\varrho$ and we find
\begin{equation}\label{pair7}
    F \sim \frac{LL_\star}{b^2}
    \exp\left[ -\tau S\left(\frac{\ln(b/L)}{\tau}\right)\right].
\end{equation}
Note that the function (\ref{pair7}) diminishes monotonically as $b$ increases. Indeed, the derivative
\begin{equation}
\frac{\partial \ln F}{\partial \ln(b/L)}
=-2-S'\left(\frac{\ln(b/L)}{\tau}\right),
\nonumber
\end{equation}
is negative since $S'(\xi)>-1$ for $\xi>0$.

The derivation above, as mentioned earlier, implies the inequality $b\gg L_-$. It is correct if $b$ is much larger than the initial value of $L_-$ (\ref{initialv}) that is $b\gg L_\star$. However, if $r_\kappa\ll b\ll L_\star$ then the optimal value of $L_-$ is $L_-\sim b$ that violates condition of the sharp peak in averaging over $\phi$ statistics, so latter just produces a factor of order unity. Since $L_-\sim b \gg r_\kappa$, the diffusion term in Eq. (\ref{lminus}) is irrelevant and we come to:
\begin{equation}
L_+=L L_\star/L_- \sim L L_\star/b.
\nonumber
\end{equation}
Therefore, providing times are large enough,  $\lambda  \tau >\ln (L_\star/b)$, we conclude
\begin{equation}
F \sim \exp\left[-\tau S\left(\frac{\ln(L_\star/b)}{\tau}\right)\right].
\label{pair9}
\end{equation}
Otherwise, $\lambda  \tau <\ln (L_\star/b)$ means $L_-\gg b$, which brings us to the mean square $F=1$.

Let us return to the fact that the pair correlation function does not depend on the diffusion coefficient above the Batchelor scale $r_\kappa$, see (\ref{pair4}). Hence, it is determined only by the statistics of Lagrangian trajectories. As it is demonstrated in Appendix~\ref{app:F-P}, there is a relation between the joint PDF ${\mathcal P}(\tau,\varrho,\phi)$ (\ref{homog9}) and pair correlation function $F$ (\ref{pair1}) in this region: $F\propto {\mathcal P}/b^2$. Dependency (\ref{pair7}) agrees with this relation as well as (\ref{pair6},\ref{pair9}) due to the symmetry (\ref{kramers3}).

As it was demonstrated in Ref.~\onlinecite{chertkov2007passive}, higher order correlation functions of the passive scalar $F_{2n}(\bm r_1, \dots , \bm r_{2n})$ in the Batchelor regime have sharp maxima in collinear geometry where the points $\bm r_1, \dots , \bm r_{2n}$ are separated in pairs with parallel differences. Let us consider such collinear geometry. It corresponds to the following leading contribution to the $2n$-th correlation function
\begin{eqnarray}
F_{2n}=\langle {\mathcal F}(b_1,\psi)
\dots {\mathcal F}(b_n,\psi) \rangle,
\label{pair10}
\end{eqnarray}
where ${\mathcal F}$ are determined by Eqs. (\ref{eq:02},\ref{factor},\ref{pair3}) and the angular brackets mean averaging over the statistics of $\phi, L_+,L_-$.

For definiteness, we consider the case $b\gg L$ where $b^2=b_1^2+ \dots +b_n^2$. Then we obtain
\begin{equation}
F_{2n}\propto \int_{\ln(b/L)}^\infty \frac{d\varrho}{b L_-^{n-1}}
\exp\left[-n\varrho -\tau S(\varrho/\tau)\right].
\label{pair11}
\end{equation}
instead of Eq. (\ref{pair5}). The integral in Eq. (\ref{pair11}) is determined by the lower limit, that is $L_+\sim b$. If $b\ll L L_\star/r_\kappa$ then we obtain the same proportionality law as in Eq. (\ref{pair7}) since $L_+ L_-=\mathrm{const}$ at the condition. Otherwise, at $b\gg L L_\star/r_\kappa$, one has $L_-\sim r_\kappa$ and
\begin{equation}
F_{2n} \propto b^{-1-n}
\exp\left[ -\tau S\left(\frac{\ln(b/L)}{\tau}\right)\right].
\label{pair12}
\end{equation}

\section{Continuous forcing of scalar}
\label{sec:continuous}

In this Section we consider the problem where fluctuations of passive scalar field $\vartheta$ are entered into the system for a long time via external random supply $f$, see Eq. (\ref{eq:25}). Then the passive scalar can be represented as an aggregation of blobs that were brought into the system by supply at different time moments and evolved thereafter. This aggregation is a sum of a big number of statistically independent contributions if the correlation time of $f$ is shorter than the characteristic time of the passive scalar evolution. Then the statistics of the passive scalar is Gaussian, as a consequence of the central limit theorem. Of course, it is valid before averaging over the flow statistics, as it was for the decaying case, see Section \ref{sec:decay}.

By analogy with the decaying case we introduce the object ${\mathcal F}(t,{\bm r}_1,\bm r_2)$ that is the product $\vartheta(t,\bm r_1) \vartheta(t,\bm r_2)$ averaged over the statistics of the forcing $f$ at a given random velocity $\bm u$. Since the statistics of $\vartheta$ is Gaussian (before averaging over the statistics of $\bm u$), any product $\vartheta(t,\bm r_1) \dots \vartheta(t,\bm r_{2n})$ averaged over the statistics of $f$ is expressed via the products of $n$ factors ${\mathcal F}$ in accordance with Wick theorem \cite{wick1950theorem}.

For definiteness, we assume that the forcing $f$ is short correlated in time. Then its statistics is determined by the pair correlation function
\begin{equation}
\langle f(t_1,\bm r_1) f(t_1, \bm r_2)\rangle
=\delta(t_1-t_2) \Theta(\bm r_1-\bm r_2).
\label{pumpcor}
\end{equation}
The expression (\ref{pumpcor}) implies homogeneity of $f$ statistics in space and time. We assume that $\Theta$ has the characteristic scale $L$ much smaller than the correlation length of the flow. However, $L$ is assumed to be much larger than the diffusion length $r_\kappa$ (\ref{rkappa}).

In our case ${\mathcal F}(t,{\bm r}_1,\bm r_2)$ is a function solely of the difference $\bm r=\bm r_1-\bm r_2$, as a consequence of spatial homogeneity of the forcing statistics. One finds from Eqs.  (\ref{eq:25},\ref{pumpcor}):
\begin{equation}
\partial_t {\mathcal F} +\Sigma y \partial_x {\mathcal F}
+ x(\partial_x u_y) \partial_y {\mathcal F}
=2\kappa \nabla^2{\mathcal F}+\Theta(\bm r).
\label{pairde2}
\end{equation}
Here, as previously, we have kept the only relevant gradient of the random velocity $\bm u$, $\partial_x u_y$. One can also substitute $\nabla^2\to \partial_y^2$ in Eq. (\ref{pairde2}).

The solution of (\ref{pairde2}) can be carried out from previous calculations for the decay problem in Section \ref{sec:decay}. Indeed, one can take solution (\ref{pairde}) with the initial condition ${\mathcal F}(t,\bm r)=\Theta(\bm r)$ at a preceeding time moment and consider its evolution till current time - so we get the contribution from one time moment of forcing $t$, the result will be obtained by integration over time interval of $f$ activity. In sense of our model, each blob brought into system has been evolving from that moment in flow and ${\mathcal F}$ is a cumulative result of all blobs over all time till now, while they could have been brought.

As for the decay case, it is instructive to analyze the Gaussian profile of the forcing correlation function $\Theta\propto \exp(-r^2/L^2)$. Then we find the solution of Eq. (\ref{pairde2}) in the large time limit:
\begin{equation}
{\mathcal F}(\bm r)
=
\int_0^\infty d\tau\, \frac{LL_\star}{L_+L_-}
\exp\left(-\Lambda_{\alpha\beta}b_\alpha b_\beta\right),
\label{pairde3}
\end{equation}
where the matrix $\hat\Lambda$ is determined by Eq. (\ref{matrixp}). The quantities $\phi,L_+,L_-$ are introduced in Section \ref{sec:decay}, their statistical properties are established there as well.

\subsection{Single-point statistics}
\label{subsec:singleperm}

Here we consider moments of the passive scalar, that is the single-point means $\langle |\vartheta|^{2\alpha}\rangle$. As in the decaying case, before averaging over the flow fluctuations the passive scalar possesses Gaussian statistics. Therefore at a given flow the moment is equal to $C_\alpha ({\mathcal F})^{\alpha}$ where $C_\alpha$ is the same as in (\ref{moments}), ${\mathcal F}$ is a value of (\ref{pairde3}) at the origin, $\bm r=0$. Thus, the moment is equal to
\begin{equation}
\langle |\vartheta|^{2\alpha}\rangle
=C_\alpha \langle ({\mathcal F})^{\alpha} \rangle,
\label{mome1}
\end{equation}

At the condition $L_\star\gg r_\kappa$ the main contribution to the single-point ${\mathcal F}$ is caused by the first stage where the product $L_+ L_-$ remains constant, the contribution is proportional to the duration of the advective stage $\tau$. Taken into account the probability of the event where $L_-$ reaches $r_\kappa$, we find from Eq. (\ref{mome1})
\begin{equation}\label{mome2}
    \langle |\vartheta|^{2\alpha}\rangle
    \sim
    C_\alpha \tau^\alpha
    \exp\left[-\tau S\left(\frac{\ln(L_\star/r_\kappa)}{\tau}\right)\right].
\end{equation}
Here the parameter $\tau$ is a subject of optimization.

At moderate $\alpha$ the maximum of the expression (\ref{mome2}) is achieved where Cram\'{e}r function $S$ is minimal, that is at $\tau=\ln(L_\star/r_\kappa)/\lambda $. This obviously leads to the Gaussian single-point statistics of $\vartheta$. In the case the moments (\ref{mome1}) are expressed via the second moment as follows
\begin{eqnarray}
\langle |\vartheta|^{2\alpha}\rangle
=C_\alpha \langle \vartheta^2 \rangle^\alpha,
\label{mome3} \\
\langle \vartheta^2 \rangle
=\frac{\Theta(0) }{\lambdabar} \ln(L_\star/r_\kappa),
\label{mome4}
\end{eqnarray}
where $\lambdabar$ is the dimensional Lyapunov exponent and $\Theta(0)$ is the scalar variance production rate. The result corresponds to one established in Ref.~\onlinecite{kolokolov2012statistical}.

However, for large exponents, $\alpha \gtrsim \ln(L_\star/r_\kappa)$, the moments strongly deviate from the relation (\ref{mome3}). Optimizing the expression (\ref{mome2}) over $\tau$, one finds the condition $\gamma\tau=\alpha$ where $\gamma$ is determined as Legendre transform of $S$, see Eqs. (\ref{homog14},\ref{homog15}). If $\alpha\ll \ln(L_\star/r_\kappa)$ then $\gamma$ is small and we return to Eq. (\ref{mome3}). If $\alpha\gg \ln(L_\star/r_\kappa)$ then $\alpha=S(0) \tau$, so $\tau \gg \ln(L_\star/r_\kappa)$ as well. The regime corresponds to the exponential tail of PDF for $\vartheta$, $\exp(-\theta /\theta_0)$, where $\theta_0^2=\Theta(0)(D\Sigma^2)^{-1/3}/S(0)$.

\subsection{Correlation functions}
\label{subsec:pairCF}

Now we move on to examine correlation functions of the passive scalar $\vartheta$. We begin with the pair correlation function. As in the decay problem, the averaging over flow statistics is required: $F({\bm r}) = \langle {\mathcal F}({\bm r})\rangle$. In the limit of long-lasting supply $F({\bm r})$ is independent of time, as a consequence of homogeneity of the forcing statistics in time. Therefore, we examine $\langle {\mathcal F}({\bm r})\rangle$ where ${\mathcal F}({\bm r})$ is given by Eq. (\ref{pairde3}).

For $b\ll r_\kappa$ we return to the second moment (\ref{mome4}). Let us consider the opposite case, $r_\kappa\ll b\ll L_\star$. Then the same logic as for the second moment does work. The main contribution to $F$ is produced by the advective stage where the product $L_+ L_-$ is a constant and the exponent in Eq. (\ref{pairde3}) can be substituted by unity. The regime is finished where $L_-$ reaches $b$. The time of the process is proportional to the corresponding logarithm and we obtain expression that is independent of the $\bm b$ vector's direction:
\begin{eqnarray}
F=\frac{\Theta(0)}{\lambdabar}
\ln(L_\star/b).
\label{pade1}
\end{eqnarray}
Formulas (\ref{mome4},\ref{pade1}) are analogous to the well-known result for the isotropic case and (\ref{pade1}) describes the Batchelor cascade of passive scalar variations towards smallest scales \cite{batchelor1959small,falkovich2001particles,donzis2010batchelor} in coordinate representation. The dependence (\ref{pade1}) can be directly obtained by solving the equation for the pair correlation function $F$, as shown in Appendix~\ref{app:F-P}.

For $b\gg L_\star$, so $L_-/b\ll 1$ at any time, after angle averaging we obtain similar to (\ref{pair5}) expression:
\begin{equation}\label{pade2}
    F\sim
    \frac{L_\star}{b}
    \int_0^\infty d\tau
        \int_{\ln(b/L)}^\infty d\varrho
    \exp\left[-\varrho -\tau S(\varrho/\tau)\right].
\end{equation}
If $b\ll L$, i.e. $\ln(b/L)<0$, integral over $\tau$ is determined by $\tau\sim 1$, hence the approximation (\ref{pade2}) is, strictly speaking, incorrect. However, one can assert, that $F\sim (\Theta(0)/\lambdabar) L_\star/b$ in the region due to the averaging over angles. If $\ln(b/L)>0$ ($b\gg L$) then $\varrho>0$ in the whole region of the integration. Then the integral over $\tau$ in Eq. (\ref{pade2}) is determined by a narrow vicinity of $\tau=\varrho/\lambda $. After integration over $\tau$ the integration over $\varrho$ will be determined by the lower limit and we find $F\sim (\Theta(0)/\lambdabar)LL_\star/b^2$. The dependence is in agreement with the stationary solution of equation (\ref{F-pumping}) for the pair correlation function $F$, see Appendix~\ref{app:F-P} and in particular~(\ref{F-mathcalPb2}).

Next, we continue our examination to high-order correlation functions $F_{2n}$. They can be represented as the sum of the $n(n-1)/2$ products of ${\mathcal F}$ in accordance with Wick theorem \cite{wick1950theorem}, where each product in the sum should be averaged over the statistics of the random flow. Below we analyze a product in the sum.

In situation where all the separations between the points are much smaller than $L_\star$, the main contribution to the average of the product of $n$ multipliers ${\mathcal F}$ is related to the advective stage. Each ${\mathcal F}$ gives the factor determined by the duration of the regime where the product $L_+L_-$ remans constant and there is no suppression related to averaging over the statistics of the angle. For moderate $n$ the evolution of $L_-$ is determined by the typical processes: $\ln(L_\star/L_-)=\lambda  \tau$. Then the duration for each ${\mathcal F}$ is proportional to the same logarithm (\ref{pade1}) and average of the product of ${\mathcal F}$'s is reduced to the product of the averages. In other words, we arrive to the Gaussian statistics for moderate number $n$ of multipliers in product, where $F_{2n}$ can be expressed via $F$ in accordance with Wick theorem \cite{wick1950theorem}.

However, if $n$ is large enough, the main contribution to $F_{2n}$ is related to rare events in which $L_-$ decrease is much slower than typically, as it was for high moments of $\theta$. In this case the value of $F_{2n}$ is determined by the duration of the event and is insensitive to the logarithms. We conclude, that in this limit $F_{2n}$ coincides with the moment $\langle (\vartheta)^{2n}\rangle$, see Section \ref{subsec:singleperm}. This non-Gaussian regime implies that $n$ exceeds logarithms for each pair correlation function (\ref{pade1}) in product. Note that the regime can be realized for lower $n$ than in one-point moments case, since  logarithms for the correlation functions are smaller than $\ln(L_\star/r_\kappa)$.

In the limit where separations are much larger than $L$, the situation is more complicated. As it was demonstrated in Ref.~\onlinecite{chertkov2007passive}, higher order correlation functions of the passive scalar $F_{2n}(\bm r_1, \dots , \bm r_{2n})$ in the Batchelor regime have sharp maxima in collinear geometry where the points $\bm r_1, \dots , \bm r_{2n}$ are separated into pairs with parallel differences. Let us consider such collinear geometry. It corresponds to the average (\ref{pair10}), where now ${\mathcal F}$ is determined by Eq. (\ref{pairde3}). Averaging over the statistics of the angle $\phi$, one finds the extra factor $(L_-/b) \exp(-b^2 L_+^{-2})$, as in Eq. (\ref{pair11}), where $b^2=b_1^2+\dots +b_n^2$. The factor $\exp(-b^2 L_+^{-2})$ implies that the effective minimum value of $\varrho=\ln(L_+/L)$ is $\ln(b/L)$. The integral over $\varrho$ is gained near it minimum value. Therefore further one can substitute $L_+=b$.

As for the pair correlation function, the distribution over times, determined by the factor $\exp[-\tau S(\varrho/\tau)]$, has a peak at $\lambda \tau=\ln(b/L)$. Therefore we find after integration over times
\begin{equation}
F_{2n}\propto \frac{1}{b^{n+1}L_-^{n-1}}.
\label{pade3}
\end{equation}
Since the expression (\ref{pade3}) is determined by typical events, we can say, that $L_-=LL_\star/b$ if $\lambda \tau<\ln(L_\star/r_\kappa)$ and $L_-=r_\kappa$ otherwise. Thus, we arrive at
\begin{eqnarray}
F_{2n}\propto \left\{
\begin{array}{cc}
b^{-2}  , & L\ll b \ll LL_\star/r_\kappa, \\
b^{-n-1}, & b \gg LL_\star/r_\kappa.
\end{array} \right.
\label{pade4}
\end{eqnarray}
The expressions (\ref{pade4}) covers the pair correlation function as well. For the pair correlation function, where $n=1$, there is no differences in the regimes of Eq. (\ref{pade4}), in accordance with the above analysis.

\section{Conclusion}

In the present paper we examined statistical characteristics of the passive scalar, like its moments and correlation functions, when it is mixed by shear flow with addition of relatively weak smooth random flow. In accordance with the established criteria, we limit our consideration to the situation when the random flow is relatively weak compared to the mean flow, (\ref{DllSigma}), but is strong enough to produce the stirring which is more intense than the molecular diffusion at the scale of the forcing, see (\ref{large-Peclet-effective}). We considered both the problem of the passive scalar decay and the problem of its statistically homogeneous in time supply. As it was expected, the statistical properties of the passive scalar appear to be far from Gaussian. Therefore, study of parameters of the passive scalar distribution cannot be reduced to the analysis of the mean square and the pair correlation function and requires examination of moments and correlation functions of higher order. We have developed the technique enabling to perform the analysis. Obtained results are expressed via Cram\'{e}r function (\ref{kramers}) for the statistics of stretching in the given random flow and thus have rather general applicability. It turns out that after the proper rescaling in space the results have properties similar to characteristic ones from the isotropic case, being written in terms of Cram\'{e}r function for their random flow. For this reason, a statistical analysis of passive scalar advection provides information about the flow statistics itself. In our work we have established properties of the Cram\'{e}r function under some general assumptions. Besides, we have provided its numerical approximation and also certain analytical results in case of flow model with short correlated in time fluctuations.

\section{Acknowledgments}

This work was performed in the Laboratory \textquotedblleft Modern Hydrodynamics\textquotedblright, established in the framework of grant 075-15-2019-1893 of the Ministry of Science and Higher Education of the Russian Federation in Landau Institute for Theoretical Physics of Russian Academy of Science, and is supported by grant No. 20-12-00383 of RScF. The authors thank I.V. Kolokolov for valuable discussions.

\section*{Data AVAILABILITY}

The data that support the findings of this study are available from the corresponding author upon reasonable request.

\appendix

\section{Symmetry of PDF for the scaling factor}
\label{sec:symmetry}

In this Section we consider a consequence of the stochastic equations (\ref{homog6},\ref{homog7}) for a random variable $\zeta$ with homogeneous in time statistics that is invariant under time inversion. The symmetry means that all correlation functions of $\zeta$ are invariant under the transformation $t\to-t$. Then the relation (\ref{symme}) is valid for the probability density function $\Pi(\varrho)$.

We consider the stochastic evolution of the variables $\varrho, \phi$ on some finite time interval $(0,T)$ and denote as $\varrho$ the value of the variable $\varrho$ at the final moment of time $\tau=T$. The function $\varrho$, as a consequence of Eq. (\ref{homog6}), can be written via time integration
\begin{equation}
\varrho=\int_0^T d\tau\, (1+\zeta) \cos\phi \sin\phi,
\label{intrho}
\end{equation}
where we assumed that initially $\varrho=0$. The random process $\zeta$ is supposed to possess homogeneous in time statistics.

Using Eqs. (\ref{homog6},\ref{homog7}), one can represent $\Pi(\varrho)$ as the following path integral \cite{feynman1965path}
\begin{eqnarray}
\Pi(\varrho)=\left\langle\int {\mathcal D}\phi\, {\mathcal D}p\,
 \exp(i I) \right.
\nonumber \\
\left. \delta\left[\varrho- \int d\tau\, (1+\zeta) \cos\phi \sin\phi\right] \right\rangle
\label{symme1} \\
I=\int d\tau\, p\left(
\partial_\tau \phi+A \right),
\label{symme2} \\
A(\phi)=-\zeta \cos^2\phi +\sin^2 \phi,
\label{defaaa}
\end{eqnarray}
where $p(\tau)$ is an auxiliary field, angle field $\phi(\tau)$ with domain on real axis, thereby containing information about rotations of $\bm \ell$, and angular brackets mean averaging over statistics of $\zeta$. The integration over $p$ in Eq. (\ref{symme1}) ensures validity of Eq. (\ref{homog7}) and the $\delta$-function in Eq. (\ref{symme1}) reflects the relation (\ref{intrho}).

Let us apply the transformation $\tau\to T-\tau$, $\varrho\to -\varrho$, $\phi\to-\phi$ that is reduced to the substitution $\zeta(\tau)\to \zeta(T-\tau)$ in the effective action $I$ and in the $\delta$-function in Eq. (\ref{symme1}). For the statistics of $\zeta$, which is invariant under the time inversion, the averages over $\zeta(\tau)$ and $\zeta(T-\tau)$ coincide. Naively, one could conclude from Eq. (\ref{symme1}) that $\Pi(\varrho)=\Pi(-\varrho)$. However, caution is needed here, since integral giving the effective action $I$ should be imposed by causality. Therefore, the time inversion is not an innocuous transformation. To clarify the point we pass to time-discretized version of the integral (\ref{symme2}).

The integral (\ref{symme2}) can be written as the limit of the sum
\begin{eqnarray}
I=\sum_{n=1}^N \left[p_n (\phi_{n}-\phi_{n-1}) + \varepsilon p_n A_{n-1}\right],
\label{symme3} \\
A_n=  - \zeta_n \cos^2 \phi_n + \sin^2\phi_n,
\label{symme4}
\end{eqnarray}
where $\varepsilon$ is the time spacing and the parameters $p_n,\phi_n$ correspond to the values of the variables $p,\phi$ at the time $\tau_n=n \varepsilon$. The value of $\phi_0$ is fixed as initial condition. In accordance with causality, the factor at $p_n$ in Eq. (\ref{symme3}) corresponds to the relation $\phi_{n}=\phi_{n-1}+\varepsilon A_{n-1}$.

Thanks to the retarded structure of (\ref{symme3}) the normalization constant
\begin{equation}
\int Dp\, D\phi\, \exp(iI)
\to \prod_{n=1}^N \int \frac{dp_n \, d\phi_n}{2\pi} \exp(iI)
\label{norma}
\end{equation}
is equal to unity. To prove this property, we begin the calculation of the integral (\ref{norma}) ``from the end'', performing first integration over the final angle, $\phi_N$, since the initial value of $\phi$, $\phi_0$, is fixed, whereas one should. The only term in $I$ (\ref{symme3}) containing $\phi_N$ is $p_N \phi_N$. Thus, the integration over $\phi_N$ produces $2\pi \delta(p_N)$ and the subsequent integration over $p_N$ is reduced to the substitution $p_N=0$. After the integrations we return to the initial form of $I$ with the number of $p_n, \phi_n$ decreased by one. Repeating the procedure, we conclude that the normalization constant equals one.

Now we consider $\Pi(-\varrho)$, that can be found by the inversion $\tau\to T-\tau$, $\phi\to-\phi$ in Eq. (\ref{symme1}). Then we arrive at the same path integral of same form, where the effective action $I$ is substituted by $I_-$. Structure of the action $I_-$ is analogous to Eq. (\ref{symme3}), though shifted, with the fixed final value $\phi_{N+1}$ now:
\begin{equation}
	\begin{gathered}
		I_-=\sum_{m=1}^N \left[p_m (\phi_{m+1}-\phi_{m}) + \varepsilon p_m A_{m}\right]\rightarrow
		\\
		\sum_{m=1}^N p_m \left[ \left(1-\varepsilon\frac{\partial A_m}{\partial \phi_m}\right)\left(\phi_{m+1}-\phi_m\right)+ \varepsilon A_{m+1}\right].
	\end{gathered}
\label{symme5}
\end{equation}
This feature leads to an additional factor at integration with the weight $\exp(iI_-)$ comparing with $\exp(iI)$ integration's weight.

To establish the factor, we consider the integral
\begin{equation}
{\mathcal N}=\int Dp\, D\phi\, \exp(iI_-)
\to \prod_{n=1}^N \int \frac{dp_n \, d\phi_n}{2\pi} \exp(iI_-).
\label{norma2}
\end{equation}
In this case, we should start ``from the beginning'', since $\phi_{N+1}$ is fixed - the first integration should be performed over $\phi_1$ and gives:
\begin{equation}
\int \frac{d\phi_1}{2\pi} \exp \left[-i p_1 \left(1-\varepsilon\frac{\partial A_1}{\partial \phi_1}\right) \phi_1 \right]
=\frac{\delta(p_1)}{\left(1-\varepsilon \frac{\partial A_1}{\partial \phi_1}\right)}.
\nonumber
\end{equation}
The subsequent integration over $p_1$ is reduced to substituting $p_1=0$.

Repeating the procedure for all $\phi_n,p_n$, one obtains
\begin{eqnarray}
{\mathcal N}=\int Dp\, D\phi\, \exp(iI_-)
\to \prod_{m=1}^N \left(1-\varepsilon \frac{\partial A_m}{\partial \phi_m}\right)^{-1}
\nonumber \\
\to \exp\left(\sum_m\varepsilon \frac{\partial A_m}{\partial \phi_m}\right)
\to \exp\left( \int d\tau\, \frac{\partial A}{\partial\phi}\right).
\nonumber
\end{eqnarray}
Thanks to the $\delta$-function in (\ref{symme1}), it is expressed via (\ref{intrho}) after our inversion transformation
\begin{equation}
	\rho(T)-\rho(0)=\frac{1}{2}\int\limits_0^Td\tau \frac{\partial A}{\partial \phi}=-\rho,
\nonumber
\end{equation}
resulting in normalization factor:
\begin{equation}
{\mathcal N}=\exp(-2\varrho).
\label{norma7}
\end{equation}
Just presence of this factor distinguishes the path integral with the effective action $I$ and one with the effective action $I_-$. Remembering, that the path integral with $I$ determines $\Pi(\varrho)$ and that the path integral with $I_-$ determines $\Pi(-\varrho)$, we arrive at the law (\ref{symme}).

\section{Connection to the unstretched space}
\label{app:rho-psi}

In the works \cite{turitsyn2007polymer,puliafito2005numerical} the other (``natural'') parametrization of vector ${\bm \ell}$,
\begin{equation}
    \ell_x = l_0 \exp\rho \cos\varphi,
    \qquad
    \ell_y = l_0 \exp\rho \sin\varphi,
\label{param-old}
\end{equation}
was used, which differs from our parametrization, see Section~\ref{subsec:lagrange}. The ``natural'' parametrization (\ref{param-old}) may be more convenient for analysis of experimental data, see e.g., Ref.~\onlinecite{liu2010stretching}. All analytical results regarding the angle dynamics and the Lyapunov exponent were obtained in Refs.~\onlinecite{turitsyn2007polymer,puliafito2005numerical} also using that parametrization. Here we demonstrate how these results can be transferred to the parametrization (\ref{homog5}) used in our work. For brevity, we introduce below the notation $\varphi_\ast = (D/\Sigma)^{1/3}$.

Angles $\phi$ and $\varphi$ are related to each other by
\begin{eqnarray}
\nonumber
    \sin\phi
    &=&
    \frac{\sin\varphi}{\sqrt{ \left(\varphi_\ast \cos\varphi\right)^2 + \sin^2{\varphi}}},
    \\
    \cos\phi
    &=&
    \frac{\varphi_\ast \cos\varphi}{\sqrt{\left(\varphi_\ast \cos\varphi\right)^2 + \sin^2{\varphi}}}.
\label{varphi-psi}
\end{eqnarray}
Under the transformation (\ref{varphi-psi}), points $\pi n/2$ (where $n$ is integer) remain unchanged. In particular, this means that the tumbling frequency $\omega$ (\ref{tumbling}) in terms of $\varphi$ and $\phi$ is the same. The exponents $\rho$ and $\varrho$ differ on a function which value is bounded in time,
\begin{equation}
    \rho=
    \varrho
    +
    \frac{1}{2}\ln\left(\frac{\cos^2\phi}{\varphi_\ast^2}+\sin^2\phi\right),
\end{equation}
so the Cram\'{e}r function (\ref{kramers}) and the Lyapunov exponent (\ref{lambda}) are the same for PDF of $\rho$. Dynamical equations (\ref{homog6},\ref{homog7}) written in terms of $\rho,\varphi$ are
\begin{equation}
    \partial_t \varphi
    =
    -\Sigma \sin^2\varphi + \partial_x u_y
    ,\qquad
    \partial_t \rho
    =
    \frac{\Sigma}{2}\sin(2\varphi)
\label{rho-psi-equation}
\end{equation}
in the limit $\varphi_\ast \ll 1$. Within the Langevin model (\ref{homog4}), noise in Eq. (\ref{rho-psi-equation}) has the statistics determined by the pair correlation function $\langle \partial_x u_y(t) \partial_x u_y(t^\prime)\rangle =2D\delta(t-t^\prime)$.

Equations (\ref{rho-psi-equation}) were obtained and treated analytically in \cite{turitsyn2007polymer,puliafito2005numerical} in the limit of small parameter $D/\Sigma\ll1$. In particular, analytical expressions for $\lambda $ and $\omega$ (\ref{lamom}) can be found there. Also, there is the stationary solution for PDF ${Q}_s(\varphi)$, which is an analogue of $P_s(\phi)$, with a narrow peak at $\varphi\sim\varphi_\ast$ of the same width $\sim\varphi_\ast$ and the algebraic tails ${Q}_s = \omega \varphi_\ast/\pi\sin^2\varphi$.

\section{Connection between the statistics of Lagrangian trajectories and the pair correlation function}
\label{app:F-P}

In this Section we establish the equation for the pair correlation function $F(t,{\bm r})$ in case of short correlated model determined by Eq. (\ref{homog4}). The function $F$ can be related to the joint PDF ${\cal P}(\phi,\varrho)$.

We start with the decay problem. The equation (\ref{thetadim}) averaged over statistics of the random flow (\ref{homog4}) takes the form
\begin{equation}
    \partial_\tau F+y \partial_w F
    -
    w^2\partial_y^2 F
    =
    r_\kappa^2 \partial_y^2 F.
\label{F-decay}
\end{equation}
One can neglect the diffusion at scales much larger than $r_\kappa$. The absence of diffusion means that the pair correlation function is determined only by the statistics of Lagrangian trajectories. In other words, equation (\ref{F-decay}) should be equivalent to equation (\ref{homog9}). To prove the property, we introduce
\begin{equation}\label{F-mathcalPb2}
    F = {\mathcal P}/b^2,
\end{equation}
where $b$ and $\psi$ are defined in (\ref{pair2}). Then the equation (\ref{F-decay}) written in terms of ${\mathcal P}$, $\varrho=\ln b$ and $\phi=\psi$ coincides with the equation (\ref{homog9}).

Next we turn to the problem of continuous forcing. The pair correlation function satisfies the same equation (\ref{F-decay}) with additional term $\Theta({\bm r})$, which is the spacial correlation function of the forcing, describing the supply, see Eq. (\ref{pumpcor}):
\begin{equation}
    \partial_\tau F+y \partial_w F
    -
    w^2\partial_y^2 F
    =
    r_\kappa^2 \partial_y^2 F
    +
    \Theta.
\label{F-pumping}
\end{equation}

To consider the Batchelor (downscale) cascade of the passive scalar, it is instructive to rewrite equation (\ref{F-pumping}) in Fourier space,
\begin{equation}
    \big(\partial_\tau
        -
        k_w\partial_{k_y}
        -
        k_y^2\partial_{k_w}^2
        + r_\kappa^2 k_y^2 \big)
    \widetilde F_{\bf k}
    =
    \widetilde \Theta_{\bf k},
\label{eq:27k}
\end{equation}
where Fourier transform is determined in accordance with
\begin{equation}
    \widetilde F_{\bf k}(t)
    =
    \int \mathrm{d}w\mathrm{d}y \, F(t,{\bm r}) \exp(-ik_ww-ik_yy).
\end{equation}
We consider (\ref{eq:27k}) in the inertial range $1/L_\star\ll k=\sqrt{k_w^2+k_y^2} \ll 1/r_\kappa$, where diffusion and supply terms are negligible. In case of the shear flow, there is a symmetry with the equation in coordinate space: in absence of diffusion, equation (\ref{eq:27k}) is equivalent to equation (\ref{F-pumping}) under change $\{k_w,k_y\}\to\{y,-w\}$. Therefore, there is the stationary solution of (\ref{eq:27k}) in the inertial range, that corresponds to the Batchelor cascade
\begin{equation}
    \widetilde F_{\bf k}
    =
    \frac{2\pi^2 \Theta(0)\,P_{s}(\phi_k)}{\lambdabar k^2},
\label{eq:40}
\end{equation}
where $k_y=k\cos\phi_k$, $k_w=-k\sin\phi_k$ and $P_s$ is the stationary angle PDF (\ref{homog9}). The factor with forcing $\Theta$ here comes from the requirement that the flux through line $k=\mathrm{const}$ should be equal to the scalar variance production rate $\Theta(0)$. The spectrum is obtained from (\ref{eq:40}) via integration over angle is $k/\left(2\pi\right)^2\int \mathrm{d}\psi\, \widetilde F_{\bf k} =  \Theta(0)/(\lambdabar k)$. The $k^{-1}$-dependence is the same as for isotropic turbulence case~\cite{donzis2010batchelor} and corresponds to logarithmic dependence (\ref{pade1}) in $b$-space. The inverse Fourier transform of (\ref{eq:40}) brings to (\ref{pade1}) in the main text.

\section{Fourier transform}
\label{sec:fourier}

Here we analyze properties of Fourier transform of the joint PDF ${\mathcal P}(\tau,\varrho,\phi)$ determined by Eq. (\ref{homog10}). One can write the equation (\ref{homog9}) in the form:
\begin{eqnarray}
\partial_\tau \widetilde{\mathcal P}
= -\hat M \widetilde {\mathcal P},
\label{homog11} \\
\hat M= (\eta -1) \cos\phi \sin\phi -\sin\phi\, \partial_\phi \sin \phi
\nonumber \\
-\left[ \cos\phi \sin\phi\, (\eta -1) +\cos\phi\, \partial_\phi \cos\phi \right]^2.
\label{homog12}
\end{eqnarray}
Thus, we arrive at the differential equation formulated solely in terms of the angle $\phi$.

A general solution of Eq. (\ref{homog11}) can be written as:
\begin{equation}
\widetilde {\mathcal P}(\tau,\phi,\eta)
=\sum_n c_n
\exp(-\gamma_n(\eta) \tau) \widetilde{\mathcal P}_n(\phi,\eta ),
\label{homog13}
\end{equation}
where factors $c_n$ depend on initial conditions. Here $\widetilde {\mathcal P}_n$ are eigenfunctions of the operator $\hat M(\eta)$ with the corresponding eigenvalue $\gamma_n(\eta)$:
\begin{equation}
\hat M \widetilde {\mathcal P}_n = \gamma_n \widetilde {\mathcal P}_n.
\nonumber
\end{equation}
The discreteness of the eigenfunctions $\widetilde {\mathcal P}_n$ is caused by their periodicity in $\phi$.

If $\gamma_n$ is the eigenvalue of the operator $\hat M$ for a given value of $\eta $, there is the same eigenvalue $\gamma_n$ for $2-\eta $. To prove the assertion we consider the matrix
\begin{equation}
M_{mn}=\int\limits_{-\pi/2}^{\pi/2} \frac{d\phi}{\pi} \exp(-2im\phi)
\hat M \exp(2in\phi),
\label{alit3}
\end{equation}
which eigenvalues are $\gamma_n(\eta)$ by definition. Let us consider the transposed matrix $M_{nm}$. Substituting $\phi\to-\phi$ in integral (\ref{alit3}) for $M_{nm}$ and integrating over $\phi$ by parts, we find $M_{nm}(\eta )=M_{mn}(2-\eta )$. Since the eigenvalues of a matrix and of a transposed one are the same, we find that eigenvalue sets of the operators $\hat M(\eta )$ and $\hat M(2-\eta )$ coincide, which concludes the assertion's proof.

At large times, the main contribution to the sum (\ref{homog13}) at a given $\eta$ is produced by the term with $\gamma_n$, which real part is the smallest (we denote it as $\gamma$). In fact, eigenvalues $\gamma_n$ are ordered by increase of $\mathrm{Re}\,\gamma_n$. Since
\begin{equation}
\widetilde \Pi=\int\limits_{-\pi/2}^{+\pi/2}\frac{d\phi }{\pi}
\widetilde {\mathcal P}(\tau,\phi,\eta),
\nonumber
\end{equation}
we conclude that just one term with this $\gamma$ enters Eq. (\ref{fourier2}). This fact enables one to find $\gamma(\eta )$ numerically, by solving the equation $\hat M\widetilde {\mathcal P}_n = \gamma_n \widetilde {\mathcal P}_n$.

\begin{figure}[t]
\includegraphics{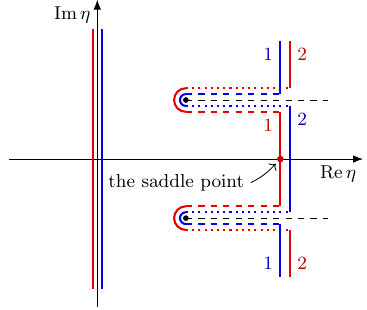}
\caption{Integration contours for the inverse Fourier transform. Continuous blue and red straight lines along the imaginary axis are the initial contours of the integration. Dashed black lines are branch cuts.}
\label{fig:invFourier}
\end{figure}

There are singular points in $\eta$ complex plane, where a pair of eigenvalue branches, $\gamma_i(\eta)$ and $\gamma_j(\eta)$ approaching to same value. In particular, the 0th eigenvalue coincides with the next 1st eigenvalue at points $\eta-1\approx \pm2.1 \pm 1.6i$, and the next two, 1st and 2nd eigenvalues with the smallest real parts are equal to each other in the points $\eta-1\approx\pm 5.0$. In the vicinity of such point, each of the corresponding eigenvalues acts as a one of two branches of a square-root-like singularity. It is important that the singularities produce zero contributions to the inverse Fourier transform
\begin{equation}
{\mathcal P}=
\int \frac{d\eta}{2\pi i}
\exp(\varrho \eta) \widetilde {\mathcal P},
\label{inversef}
\end{equation}
 at large $|\varrho|$. Let us take negative $\varrho$, then one should deform two contours of integral which correspond to the involved eigenvalues as depicted in Figure~\ref{fig:invFourier}. The integrands including pre-exponents, considered as analytical continuation from the real axis, are equal to each other on the dotted parts of the deformed contours, but the directions of contours are opposite, so the results of integrations along the contours' dotted parts cancel each other. The same is for the dashed parts of the contours. Because of branch points' square-root-like behavior, one can reassemble the integration contours after passing the point into new pair of contours from the pieces, which are marked as \textquoteleft1\textquoteright\ and \textquoteleft2\textquoteright\ in Fig.~\ref{fig:invFourier}. Therefore, we conclude that moving integration contours in the pairs on the complex plane allows one not consider such singular branch points.

In that way, at large times the inverse Fourier transform is determined by the 0th eigenvalue and eigenfunction and the corresponding saddle point located on the real axis, that ensures the positiveness of the PDF ${\mathcal P}$. The saddle point satisfies the equation (\ref{homog15}): $\eta=-\partial_\xi S(\xi)$. Knowing $\gamma(\eta)$, one finds the Cram\'{e}r function via the Legendre transform. The result is in consistence with the statistics obtained via numerical simulation of Langevin equations Eq. (\ref{homog6},\ref{homog7}). Note that the Cram\'{e}r function $S(\xi)\approx 0.33 \xi^4$ at large $|\xi|\gg1$. This extremely fast decay of the PDF stops at $|\xi|\sim (\Sigma/D)^{2/3}$, after that the Cram\'{e}r function $S\sim (\varrho/Dt)^2$ that is provided by terms with the derivatives of $\bm u$ in Eqs. (\ref{elldynamics1},\ref{elldynamics2}), omitted in our analysis.


\end{document}